\documentclass[journal,twoside,draftversion]{ieeecolor}
\usepackage{hyperref}
\hypersetup{pdfcreator={Dario Sitnik}}
\usepackage[T1]{fontenc}
\usepackage{tmi}
\usepackage{cite}
\usepackage{amsmath,amssymb,amsfonts}
\usepackage{algorithmic}
\usepackage{multirow}
\usepackage{float}
\usepackage{caption}
\usepackage{subcaption}
\usepackage{graphicx}
\usepackage{textcomp}
\def\BibTeX{{\rm B\kern-.05em{\sc i\kern-.025em b}\kern-.08em
    T\kern-.1667em\lower.7ex\hbox{E}\kern-.125emX}}

\begin{document}
\title{LEFM-Nets: Learnable Explicit Feature Map Deep Networks for Segmentation of Histopathological Images of Frozen Sections}

\author{
    Dario Sitnik and Ivica Kopriva\textsuperscript{*}, Senior Member, IEEE
}
\maketitle

\newcommand\nnfootnote[1]{%
  \begin{NoHyper}
  \renewcommand\thefootnote{}\footnote{#1}%
  \addtocounter{footnote}{-1}%
  \end{NoHyper}
}

\nnfootnote{This work has been supported in part by the Croatian Science Foundation Grant IP-2016-06-5235, and in part by the European Regional Development Fund under the grant KK.01.1.1.01.0009 (DATACROSS).}
\nnfootnote{Dario Sitnik is with the Division of Electronics, Ruđer Bošković Institute, Bijenička cesta 54, Zagreb, Croatia. e-mail: dario.sitnik@irb.hr.}
\nnfootnote{Ivica Kopriva is with the Division of Electronics, Ruđer Bošković Institute, Bijenička cesta 54, Zagreb, Croatia. e-mail: ikopriva@irb.hr.}
\begingroup\renewcommand\thefootnote{*}
\footnotetext{\textit{Corresponding author}}

\begin{abstract}\hbadness=2500
Accurate segmentation of medical images is essential for diagnosis and treatment of diseases. These problems are solved by highly complex models, such as deep networks (DN), requiring a large amount of labeled data for training. Thereby, many DNs possess task- or imaging modality specific architectures with a decision-making process that is often hard to explain and interpret. Here, we propose a framework that embeds existing DNs into a low-dimensional subspace induced by the learnable explicit feature map (LEFM) layer. Compared to the existing DN, the framework adds one hyperparameter and only modestly increase the number of learnable parameters. The method is aimed at, but not limited to, segmentation of low-dimensional medical images, such as color histopathological images of stained frozen sections. Since features in the LEFM layer are polynomial functions of the original features, proposed LEFM-Nets contribute to the interpretability of network decisions. In this work, we combined LEFM with the known networks: DeepLabv3+, UNet, UNet++ and MA-net. New LEFM-Nets are applied to the segmentation of adenocarcinoma of a colon in a liver from images of hematoxylin and eosin (H\&E) stained frozen sections. LEFM-Nets are also tested on nuclei segmentation from images of H\&E stained frozen sections of ten human organs. On the first problem, LEFM-Nets achieved statistically significant performance improvement in terms of micro balanced accuracy and $\mathbf{F}_{1}$ score than original networks. When averaged over ten runs, LEFM-MA-net achieved balanced accuracy of $\mathbf{89.36 \boldsymbol{\%} \boldsymbol{\pm} 1.28 \boldsymbol{\%}}$ compared to $\mathbf{88.02 \boldsymbol{\%} \boldsymbol{\pm} 1.22 \boldsymbol{\%}}$ by the MA-net. Corresponding results for $\mathbf{F}_{1}$ score are $\mathbf{84.96 \boldsymbol{\%} \boldsymbol{\pm} 1.14 \boldsymbol{\%}}$ and $\mathbf{82.75 \boldsymbol{\%} \boldsymbol{\pm} 1.10 \boldsymbol{\%}}$. On the second problem, LEFM-Nets achieved only better performance in comparison with the original networks. LEFM-MA-net achieved balanced accuracy of $ \mathbf{89.41 \boldsymbol{\%} \boldsymbol{\pm} 0.29 \boldsymbol{\%}}$ compared to $\mathbf{89.30 \boldsymbol{\%} \boldsymbol{\pm} 0.44 \boldsymbol{\%}}$ by the original MA-net. Results for $\mathbf{F}_{1}$ score are $\mathbf{85.35 \boldsymbol{\%} \boldsymbol{\pm} 0.25 \%}$ and $\mathbf{85.12 \% \boldsymbol{\pm} 0.51 \boldsymbol{\%}}$. The source code is available at \textit{https://github.com/dsitnik/lefm}
\end{abstract}

\hfill
\section{Introduction}
\label{sec:introduction}
\IEEEPARstart{S}{egmentation} and/or classification of medical images plays a vital role in disease diagnosis, therapy planning and follow-up tracking of therapy efficiency. Therefore, segmentation/classification by using deep learning (DL) models, i.e. deep neural networks (NNs), is steadily in focus in biomedical image analysis \cite{shen2017deep,litjens2016deep,de2018clinically,mishra2022data,gu2021net,gunesli2020attentionboost,li2021pathal,yang2020guided,zhu2021hard,guo2021learn}, to cite a few. Thus, a constant need for performance improvement results in many specialized and extremely complex DL models, whereat some of them are cited above. Their decision-making process is often hard to explain and sometimes even harder to interpret in terms of features understandable to medical experts \cite{rudin2019stop,tjoa2020survey}. When high-stakes judgments must be made, it is critical to comprehend the reasons behind the decision-making process.
As a result, rather than explaining black box models, it is preferable to create interpretable models  \cite{rudin2019stop}. Furthermore, there is currently a scarcity of expert-annotated datasets in many medical imaging modalities. For sophisticated models like convolutional NNs  (CNNs), this causes a generalization challenge  \cite{litjens2016deep,veta2014breast,qiu2018deep}. This issue is also present in histopathological image segmentation of intraoperatively collected frozen sections \cite{komura2018machine}. One reason is that sample collection during surgery is distracting. Also, another reason is highly time-consuming effort for pathologists to label cancerous pixels \cite{sitnik2021cocahis}. 

\begin{figure*}[ht]
\centering
\includegraphics[width=0.9\textwidth]{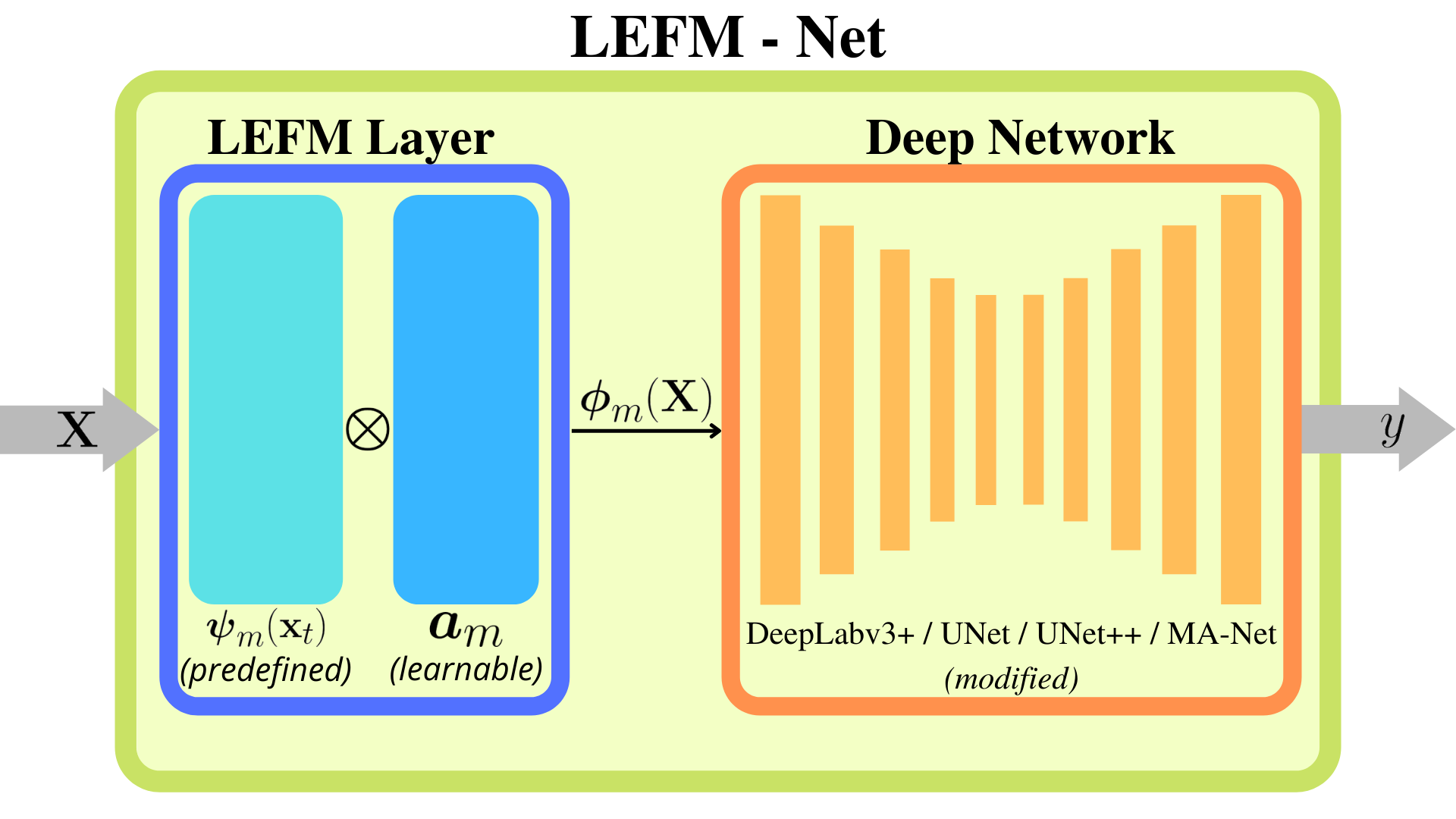}
\caption{Proposed framework comprising the slightly modified existing deep networks such as UNet, UNet++, DeepLabv3+, and MA-net, embedded in a space induced by the learnable explicit feature map (LEFM) layer. Increased number of learnable weights in the LEFM layer is represented as $\mathbf{a}_{m}$. Additional deep network's weights are introduced through modification to accept expanded features. $\phi_{m}(\mathbf{X})$ is defined as Hadamard product ($\otimes$) of: $\psi_{m}(\mathbf{x_t})$ - the spatial location dependent map of monomials of the input data space features, and $\mathbf{a}_{m}$ - the spatial location invariant vector of coefficients. The first part contributes prior knowledge in terms of monomial-based predefined algebraic structure. The second part learns the importance of explicit features.}
\label{figure:adafel} 
\end{figure*}

The issues outlined above motivated us to propose an inclusive framework for DL-based segmentation/classification of medical images in general, and color histopathological images of frozen sections in particular. Instead of designing specialized DL models, we aim to improve segmentation/classification performance of the existing NNs, such as UNet \cite{ronneberger2015u}, UNet++ \cite{zhou2018unet++,zhou2020unet++}, DeepLabv3+ \cite{chen2018encoder} and MA-net \cite{fan2020manet}. To achieve that goal, we embed the existing network in a space induced by the learnable explicit feature map (LEFM) layer (Fig \ref{figure:adafel}). The LEFM layer is an analog to learnable kernels \cite{wang2021bridging,pan2011learning,kulis2009low,li2015adaptive,bach2004multiple,rebai2016deep,wilson2016deep,bohn2019representer}, and its use is motivated by the need to create low-dimensional subspace in data-adaptable Hilbert space where existing network will yield improved performance. To reduce risk of overfitting, the aim was to only modestly increase the complexity relative to the original network. That is achieved by using the LEFM of low order, whereat the LEFM order is the only new hyperparameter introduced by the proposed framework. Its suggested value is either 2 or 3, which simplifies the cross-validation procedure.

The explanation why the LEFM layer improves performance of deep network is in representation of the LEFM. The LEFM is represented as Hadamard (entry-wise) product of the spatial location invariant vector of coefficients and spatial location dependent map composed of the monomials of features from the input data space. Thus, the predefined (known) algebraic structure of the EFM contains \textit{a priori} knowledge that contributes to improved performance of the proposed framework. For example, \textit{a priori} knowledge in a form of translation symmetry in images is incorporated in CNNs \cite{lecun1998gradient}, and enables them to achieve translation equivariance by re-using convolutional filters at all spatial locations. Such design highly reduces the number of learnable parameters in CNNs. Analogously, the learned vector of coefficients, in proposed LEFM, can be re-used at all spatial locations with location dependent map of monomials. Learnable EFM alleviates the analogous problem in kernel-based nonlinear methods: the choice of kernel. After learning, the coefficients emphasize the importance of the features in LEFM. Since all features are known polynomial functions of the original features from the input data space, that contributes to explainability and interpretability of the decision-making process of the proposed framework. The main contributions of this work are listed as follows:

\begin{itemize}
    \item Framework for DL-based segmentation/classification of low-dimensional medical images comprising slightly modified existing deep network embedded into a subspace induced by learnable EFM layer.
    \item Representation of learnable EFM in terms of spatial location dependent map of monomials of features from the input data space and spatial location invariant vector of coefficients. The first part brings prior knowledge in terms of predefined algebraic structure with monomial-based explicit features. The second part learns the importance of explicit features. That contributes to medically explainable and interpretabe decision-making. The overall representation modestly increases the number of learnable parameters compared to the original network.
    \item Extensive validation of the proposed framework where we used existing deep architectures: DeepLabv3+, UNet, UNet++, and MA-net on two problems. The first one was image segmentation of adenocarcinoma of a colon in a liver from hematoxylin and eosin (H\&E) stained frozen sections. The second problem was the segmentation of nuclei from H\&E stained frozen section images of ten human organs. Proposed framework was compared to the original networks. In the first problem, statistically significant improvements in terms of micro balanced accuracy, micro $F_{1}$ score and micro precision are obtained. In the second, the proposed framework yielded only better results.
\end{itemize}

\newpage
\section{Related work}

\subsection{Brief overview of some specialized deep neural networks}

Herein, we briefly comment on recently proposed deep networks with task- or imaging modality specific architectures with an emphasis on histopathological image analysis. New deep CNN was proposed in \cite{mishra2022data} for medical image segmentation. It exploits specific attributes in the input datasets. That is achieved with auxiliary supervision implemented in terms of object perceptive field and layer-wise effective receptive fields. Comprehensive attention-based CNN (CA-Net) is proposed in \cite{gu2021net} for more accurate and more explainable medical image segmentation. That is achieved by using multiple attention modules in CNN architecture aware of the most important spatial positions, channels and scales\cite{gu2021net}. Although the explainability of decision is improved, the interpretability of the features is still problematic. AttentionBoost network is proposed in \cite{gunesli2020attentionboost} for gland instance segmentation in histopathological images. It is based on a multi-attention learning model that, based on adaptive boosting, adjusts loss of fully convolutional networks to adaptively learn what to attend at each stage. In \cite{li2015adaptive}, a new methodology, named PathAL, is presented aiming to improve classification performance in histopathology image analysis with a reduced number of expert annotations. Guided soft attention network for classification of histopathological breast cancer images is proposed in \cite{yang2020guided}. In particular, region-level supervision is used to guide attention of CNN. The attention mechanism activates neurons in diagnostically relevant regions and suppresses activation in diagnostically irrelevant regions. Hard sample aware noise robust learning deep network for classification of histopathology images is proposed in \cite{zhu2021hard}. The method is aimed at distinguishing informative hard samples from the harmful noisy ones. That lowers the noise rate and yields an almost clean dataset for training network in a self-supervised manner. Novel ThresholdNet \cite{guo2021learn} is designed for automatic segmentation of polyp in endoscopy images such that segmentation threshold is learnable. Proposed LEFM-Net framework is directly applicable to methods proposed in \cite{mishra2022data,gu2021net,gunesli2020attentionboost,li2021pathal,yang2020guided,zhou2020unet++,guo2021learn}. In this paper, however, we demonstrated performance improvement of the LEFM-Net framework using well know deep networks: DeepLabv3+, UNet, UNet++ and MA-net.

\subsection{Brief overview of kernel-based learning}

Let us assume there is a function $f:=\mathbb{R}^{d} \rightarrow \mathbb{R}$ that assigns the class label $y \in S_{C}$ to the sample $\mathbf{x} \in \mathbb{R}^{d}$. Here, $d$ stands for dimension of the ambient input space, $C$ stands for the number of classes and $S_{C}$ is a set containing all possible labels. As an example, for binary classification problem we have $C=2$ and $S_{2}=\{0,1\}$ or $S_{2}=\{-1,1\}$. The classifier learning problem is to estimate $f$ from the given set of training data $\left\{\left(\mathbf{x}_{t}, y_{t}\right)\right\}_{t=1}^{T}$. The representer theorem \cite{scholkopf2018learning,kimeldorf1971some}, states that each function $f$ in Hilbert space $\mathcal{H}$, a.k.a. feature space, that minimizes an arbitrary loss function, admits representation:

\begin{equation}
    \label{eq:repre}
    f(\mathbf{x})=\sum_{t=1}^{T} \alpha_{t} \kappa\left(\mathbf{x}_{t}, \mathbf{x}\right) 
\end{equation}

\noindent where $\{ \alpha_t \}_{t=1}^{T}$, are estimated from the training data. That gave rise to the development of many machine learning algorithms, such as support vector machines and kernel principal component analysis, to name a few. These methods work well when the underlying problem fits $\mathcal{H}$ properly, i.e. when the given empirical values stem from $f \in \mathcal{H}$. In other words, the kernel function $\kappa(\circ, \circ)$ has to be chosen properly. If the appropriate $\mathcal{H}$ is not known, one can resort to data adaptive kernel methods \cite{pan2011learning,kulis2009low,li2015adaptive}, multiple kernel learning methods \cite{bach2004multiple}, multi-layer (deep) multiple kernel learning method \cite{rebai2016deep}, and deep kernel method \cite{wilson2016deep}. In particular, the last two approaches combine flexibility of deep networks. One where feature detection is done automatically with the approximation power of kernel methods. The second one is where the feature map is determined implicitly by the kernel function. Moreover, in \cite{bohn2019representer} the representer theorem is proven for a multilayer network of $L$ concatenated kernel functions. The power of kernel methods is their capability to implicitly learn potentially infinite dimensional nonlinear features in the reproducing kernel Hilbert space $\mathcal{H}$. The computation is made tractable by leveraging the kernel trick \cite{bohn2019representer}. However, implicitly learned features are hard to interpret. 

\section{Learnable feature maps and deep networks}

As pointed out in \cite{wilson2016deep}, one of the main reasons to exploit deep kernel learning was growing frustration among researchers with too many design options associated with deep networks. Furthermore, the decision-making process of deep networks is also difficult to interpret. However, deep networks composed of multiple layers of nonlinear functions can approximate a rich set of naturally occurring input-output dependencies \cite{wang2021bridging}. Thus, they can achieve high classification performance. Therefore, we aim to propose a framework that further improves the performance of deep networks, contributes to the interpretability of the results and brings a modest increase in the number of learnable parameters compared to original networks. Towards that goal, we propose to embed existing deep networks in a subspace induced by the LEFM of small order (typically 2 or 3). Our approach relies on the well know kernel trick \cite{scholkopf2018learning}: for any positive definite kernel function $\kappa(\mathbf{x}, \mathbf{y})$ there exists a function $\phi(\mathbf{x})$ that maps data $\mathbf{x}$ to a Hilbert space $\mathcal{H}$, such that $\kappa(\mathbf{x}, \mathbf{y})=\langle\phi(\mathbf{x}), \phi(\mathbf{y})\rangle_{\mathcal{H}} \ \cdot \  \phi(\mathbf{x})$ is known as EFM. Thus, we can recast the representer theorem (\ref{eq:repre}) in terms of kernel associated EFM:

\begin{equation}
    f(\mathbf{x})=\langle\mathbf{w}, \phi(\mathbf{x})\rangle_{\mathcal{H}}
\end{equation}

\noindent where $\mathbf{w}=\sum_{t=1}^{T} \alpha_t \phi\left(\mathbf{x}_{t}\right)$. Thus, the nonlinear approximation problem (\ref{eq:repre}) in the input data space becomes the linear one in the feature space $\mathcal{H}$. Hence, kernel-based nonlinear algorithms can be seen as linear algorithms operating in appropriate feature space \cite{vedaldi2012efficient}. That is formalized by the Cover’s theorem \cite{cover1965geometrical}. It states that the number of separating hyperplanes is proportional to the dimension of the feature space. Thus, it is important for the dimension of expanded feature space $D$ to satisfy $D>>d$. That is fulfilled easily since EFMs associated with many popular kernels are mostly high dimensional or even infinite dimensional (e.g. EFM associated with the Gaussian kernel). That is also a reason feature maps are not used explicitly in nonlinear algorithms \cite{vedaldi2012efficient}. Nevertheless, we can partially circumvent this obstacle by using the approximate EFM of order $m$, i.e. $\phi_{m}(\mathbf{x})\in\mathbb{R}^D$. In that regard, we emphasize that EFMs can be written in a common algebraic form:

\begin{equation}
\phi_{m}(\mathbf{x})=\left[\left\{a_{q_{1}, \ldots ,q_{d}} x_{1}^{q_{1}} \ldots x_{d}^{q_{d}}\right\}_{q_{1}, \ldots, q_{d}=0}^{m}\right]^{\mathrm{T}}
\end{equation}

\begin{equation*}
\text{such that} \sum_{i=1}^{d} q_{i} \leq m. 
\end{equation*}
When $m \rightarrow \infty$ we have: $\phi_{m}(\mathbf{x}) \rightarrow \phi(\mathbf{x})$. Bearing in mind that the dimension of induced feature space $D$ depends on $d$ and $m$ through: 

\begin{equation}
\label{eq:binom}
D = \binom{d+m}{m}
\end{equation}

\noindent it becomes clear that we need to keep $m$ low if we do not want to introduce too many learnable parameters in the proposed framework. Thus, in the experiments conducted in this paper, we have $m \in\{2,3\}$. Hence, the formulation (2) with approximate EFM $\phi_{m}(\mathbf{x})$ will only approximate the original nonlinear approximation problem (1). That is why we need to embed the deep network in a subspace $\mathcal{H}_{D} \subset \mathcal{H}$ induced by $\phi_{m}(\mathbf{x})$. By doing so, we expect to further boost segmentation/classification performance of the network. Nevertheless, the problem related to the proper fit of the underlying problem to $\mathcal{H}$ still exists. Thus, we propose the following computationally efficient representation of $\phi_{m}(\mathbf{x})$:

\begin{equation}
\label{eq:phim}
    \phi_{m}(\mathbf{x}):= \psi_{m}(\mathbf{x}) \otimes \mathbf{a}_{m} 
\end{equation}

\noindent where $\otimes$ denotes Hadamard (entry wise) product, and:

\begin{equation}
\label{eq:psim}
\psi_{m}(\mathbf{x}):=\left[\left\{x_{1}^{q_{1}} \ldots x_{d}^{q_{d}}\right\}_{q_{1}, \ldots, q_{d}=0}^{m}\right]^{\mathrm{T}} \text{s.t.} \sum_{i=1}^{d} q_{i} \leq m ,
\end{equation}

\begin{equation}
\label{eq:am}
\mathbf{a}_{m}:=\left[\left\{a_{q_{1}, \ldots , q_{d}}\right\}_{q_{1}, \ldots, q_{d}=0}^{m}\right]^{\mathrm{T}} \text {s.t.} \sum_{i=1}^{d} q_{i} \leq m .
\end{equation}

Hence, when existing deep network is trained with $\phi_{m}(\mathbf{x})$, the coefficients $\mathbf{a}_{m}$ are learned together with the rest of network parameters. Since $\mathbf{a}_{m}$ is spatial location invariant, learning is computationally efficient. As opposed to that, $\psi_{m}(\mathbf{x})$ is a spatial location dependent function of the input features and it is not learnable. Thus, when $d$ and $m$ are small, the number $D$ of learnable coefficients in $\mathbf{a}_{m}$ is modest when compared to the millions of parameters in existing deep networks. Number of trainable parameteres is additionally increased in amounts specific to the deep architecture and its backbone. Thus, all newly introduced parameters are directly related to the increase from $d$ to $D$ dimensional input. However, vast majority of the original network remains the same.

\noindent As an example, in the problem considered in the paper, we apply the proposed framework to segmentation of color (RGB) histopathological images of H\&E stained frozen sections. Thus, we have $d=3$. As we worked with approximate EFM of order $m \in\{2,3\}$, that respectively yields $D \in\{10,20\}$. The rationale why the deep network, embedded in a subspace $\mathcal{H}_{D} \subset \mathcal{H}$ induced by learnable EFM, outperforms the original network operating in the input data space is due to the explicit character of features up to the selected order $m$. Through learning, a deep network can determine their importance. Thus, in the proposed framework, the network will use more explicit features than it is available in the input data space. Since features in $\phi_{m}(\mathbf{x})$ are polynomials comprising the originals features from the input data space, it is possible to interpret them.
\section{Experiments and Results}
\label{sec:expRes}
\subsection{Data adaptive feature expansion layer}
\label{subsec:layer}
For implementing an LEFM layer, fundamental functionality is described in (\ref{eq:phim}) - (\ref{eq:am}). As mentioned, the coefficients are data adaptive (i.e. learnable) and, for LEFM layer construction, we are only interested in monomials. Thus, we need to find monomials (\ref{eq:psim}) using all combinations of $\boldsymbol{q}$ such that $\sum_{i=1}^d{q_i}\leq m,\ \forall q_i \geq 0$. Thus, the number of terms in this setting becomes (\ref{eq:binom}). Optionally, if the LEFM layer is followed by batch normalization layer \cite{ioffe2015batch}, two parameters per each term (mean and standard deviation) are introduced. 

The core functionality of LEFM layer is developed using two matrices, term- and power mask. Both matrices are $D\times d$ dimensional. The term mask matrix $\mathbf{TM}$ is associated with selecting features $\{x_i \}^d_{i=1}$ and the power mask matrix $\mathbf{PM}$ with exponantiating selected features by $\{q_i \}^d_{i=1}$. A resulting monomial matrix $\mathbf{MM}$ is made by element-wise exponentiation $\mathbf{MM}=\mathbf{TM}^{\mathbf{PM}}$. Thus, when the resulting matrix is squeezed by elements-product across the $2^{nd}$ dimension, we get the monomials of the input features $\psi_m(\mathbf{x})$. Using the randomly initialized vector of learnable coefficients $\mathbf{a}_{m}$ we apply the Hadamard product as in (\ref{eq:phim}).

Thus, with $d$ features as an input, the LEFM layer gives $D$ output features with learnable coefficients $\{ a_i \}^D_{i=1}$. As illustrated in Fig. \ref{figure:adafel}, the expanded output is passed to the arbitrary deep NN to jointly train its weights and LEFM layer’s coefficients. 

\begin{figure*}[ht!]
\centering
\includegraphics[width=\textwidth]{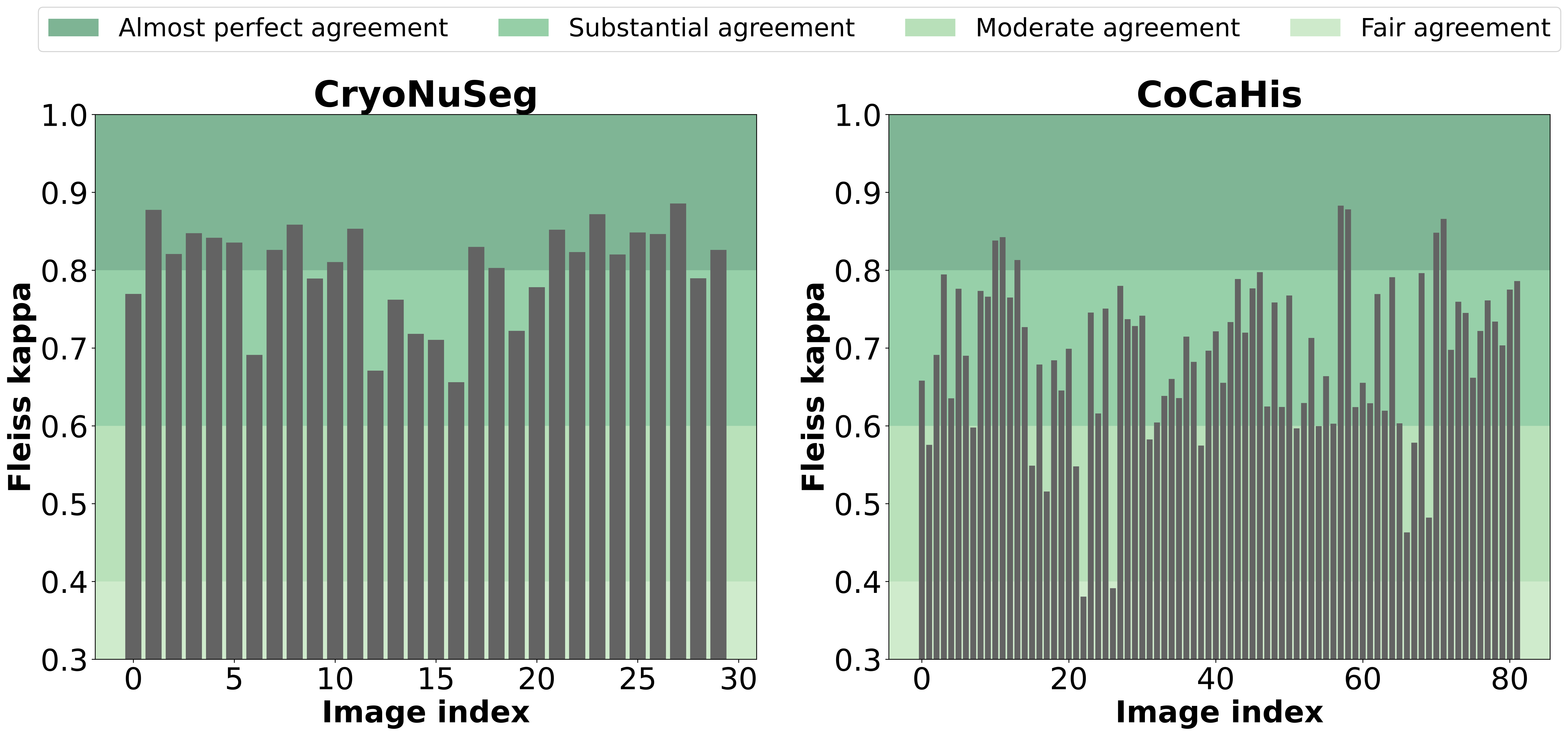}
\caption{Fleiss kappa statistics for evaluating inter-annotators agreement. Left and right plots represent statistics per image in CryoNuSeg and CoCaHis datasets, respectively. The Fleiss kappa for the entire CryoNuSeg is 0.8012 and for CoCaHis 0.7405. Former indicates an almost perfect inter-annotator agreement, whereas the latter indicates a substantial inter-annotator agreement.}
\label{figure:fleisskappa} 
\end{figure*}

\subsection{Deep neural networks}
\label{subsec:dnn}
Within the scope of this paper, the LEFM layer is combined with several types of slightly modified neural networks. In DeepLabv3+ \cite{chen2018encoder}, a deep network has the spatial pyramid pooling module combined with the encoder-decoder structure. Feature extraction at arbitrary resolution was enabled by atrous convolution layers contained in the encoder module \cite{chen2018encoder}. On the other hand, a simple decoder recovers detailed boundaries of an object. A popular UNet model \cite{ronneberger2015u} employs a symmetric U-shaped encoder-decoder for precise feature localization. Skip connections help in preserving context at the same level of the model’s depth \cite{ronneberger2015u}. Natural extension of the UNet is a nested model. Introduced in \cite{zhou2018unet++}, UNet++ model proposes additional nodes and skip connections to the original UNet. That facilitates predicting the output from different levels of depth. Training the entire network by combining all outputs, the UNet++ model outperforms UNet with a cost of a larger number of parameters. Finally, MA-net \cite{fan2020manet} incorporates a self-attention mechanism. Based on the attention mechanism, it collects rich contextual dependencies while adaptively integrating local features. Model implements Position-wise Attention Block and Multi-scale Fusion Attention Block. Prior is used to model the feature interdependencies spatially, and the former to capture channel dependencies using semantic feature fusion \cite{fan2020manet}. 

All mentioned models can implement different backbones for feature extraction to their core. When comparing a model with and without the LEFM layer, it is important to keep the backbone fixed. MobileNetV2 \cite{sandler2018mobilenetv2} was implemented in DeepLabv3+ model, whereas other models have DenseNet201 \cite{huang2017densely} backbone. All backbone’s weights are pretrained on Imagenet classification problem \cite{deng2009imagenet} and fine-tuned with the rest of the model. Thus, the total number of trainable parameters in DeepLabv3+ model is \textasciitilde4.4M, in UNet \textasciitilde28.5M, in UNet++ \textasciitilde48.5M, and in MA-net \textasciitilde133.7M. Furthermore, LEFM-Nets, with mentioned architectures and $m=3$, contribute to an increase of the number of learnable parameters in the amount of: 0.11\%, 0.18\%, 0.11\%, and 0.04\% relative to the original networks. In the application considered herein, it is modest when compared to the reported total number of trainable parameters of original models.

\subsection{Datasets}
This paper is focused on segmentation of color images of H\&E stained frozen sections. Thereby, we used the only two publicly available datasets: CoCaHis \cite{sitnik2021cocahis,cocahis2021dataset} and CryoNuSeg \cite{mahbod2021cryonuseg}. Datasets are divided into the train and test sets. Furthermore, during the training process, 20\% of the train set is left for validation and early stopping purposes.

\subsubsection{CoCaHis}
Dataset comprises 82 images with dimensions $1037 \times 1388$. Training set has 58 images belonging to 13 patients. The rest of 24 images from 6 patients are contained in the test set. Sets are disjointed for both images and patients. Seven annotators generated the pixel-wise labels for each image. One and zero indicate cancerous and non-cancerous pixels, respectively. Resulting average Fleiss kappa statistic \cite{fleiss1971measuring}, presented in Section \ref{sec:results} and Fig. \ref{figure:fleisskappa}, supports usage of majority voting due to the substantial inter-annotator agreement \cite{landis1977measurement}.

\renewcommand{\thefootnote}{\arabic{footnote}}
\subsubsection{CryoNuSeg} Collected from The Cancer Genome Atlas\footnote[1]{https://www.cancer.gov/tcga}, the dataset contains 30 whole-slide frozen section images. Afterwards, they are cropped to the most representative window. Hence, the dataset comprises 30 images of the size $515\times 512$ pixels that correspond to 10 organs. The problem formulated by this dataset is segmentation of nuclei in different images. Therefore, the ground truth generation is similar to CoCaHis. The Fleiss kappa statistic, presented in Section \ref{sec:results} and Fig. \ref{figure:fleisskappa}, suggests an almost perfect agreement between three annotations, and that is why the majority vote is used. To ensure the representation of every organ in both train and test set, one image per each organ is left out for the test set (10 images), whereas the remaining images are included in the training set (20 images).

\subsection{Statistics and Metrics}
\subsubsection{Statistical significance analysis}
To determine if inclusion of the LEFM layer yields statistically better performance when compared to the original models, we used a one-way analysis of variance (ANOVA) test. All models were trained 10 times with different initial weights and different subsets for validation. Null hypothesis was that two groups of results, obtained by the LEFM-Net and the original deep network, have equal means. For all tests, a level of significance $\alpha=0.05$ was used.

\subsubsection{Evaluation metrics}
\label{sec:metrics}
All metrics used for evaluation were calculated using standard $TP$, $TN$, $FN$, and $FP$ notation. $TP$ stands for a number of correctly diagnosed cancerous pixels. $TN$ indicates a number of  non-cancerous pixels correctly diagnosed as non-cancerous. $FN$ stands for a number of cancerous pixels incorrectly diagnosed as non-cancerous. Finally, $FP$ indicates the number of noncancerous pixels incorrectly diagnosed as cancerous. Thus, the following micro measures were used for models’ performance validation:
\begin{itemize}

\item a harmonic mean of precision and sensitivity (recall) presented as a $F_1$ score:
\begin{equation}
\label{eq:f1}
    F_1 = \frac{2\cdot TP}{2\cdot TP+FP+FN},
\end{equation}

\item balanced accuracy (BACC) as an arithmetic mean of sensitivity (SE) and specificity (SP):
\begin{equation*}
SE = \frac{TP}{TP+FN},\ SP = \frac{TN}{{TN}+{FP}},
\end{equation*}

\begin{equation}
\label{eq:bacc}
BACC = \frac{SE+SP}{2},
\end{equation}

\item and Precision score, which measures the classifier's capacity not to categorize a sample as positive if it is negative:
\begin{equation}
\label{eq:prec}
    PREC = \frac{TP}{TP+FP}.
\end{equation}
\end{itemize}

\noindent For all the metrics, 0 indicates the worst performance and 1 indicates the best performance.

\begin{figure}[H]
    \centering
    \includegraphics[width=0.35\textwidth]{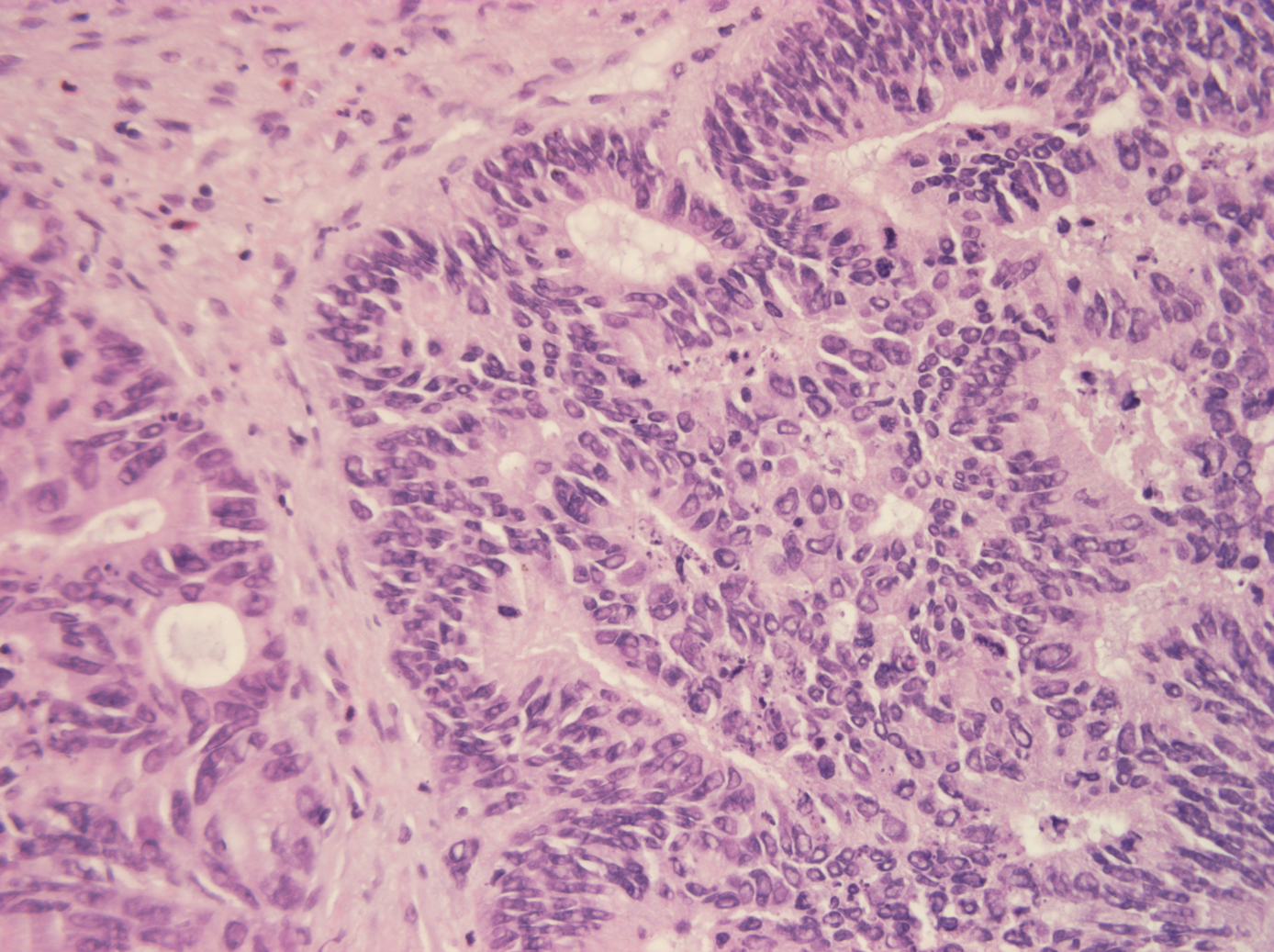}
    \caption{Target image for stain normalization chosen by pathologist. Selection is based on the quality-of-staining criterion.}
    \label{figure:target_image}
\end{figure}

\subsection{Data preprocessing}
\label{sec:preprocessing}

As an important part of the computational pathology pipeline, stain normalization was performed to eliminate the variability of staining and experimental variations. The structure preserved color normalization method \cite{vahadane2016structure} was used on both CoCaHis and CryoNuSeg datasets. A target image, shown in Fig. \ref{figure:target_image}, was carefully chosen by an experienced pathologist, ensuring a well stained specimen for the mentioned algorithm. Fig. \ref{figure:panel} shows original and stain-normalized images from the mentioned datasets. Stain-transfer from the target image helps models during the training to cope with the color variation problem.

Prior to the training, every image was scaled to the range [0,1]. Additionally, random shifting, scaling, rotating and flipping (horizontal, vertical) were applied as part of the data augmentation procedure. Chosen parameters are presented in Section \ref{sec:implDetail}. 

\begin{figure}[H]
     \centering
     \begin{subfigure}[b]{0.24\textwidth}
         \centering
         \includegraphics[width=\textwidth]{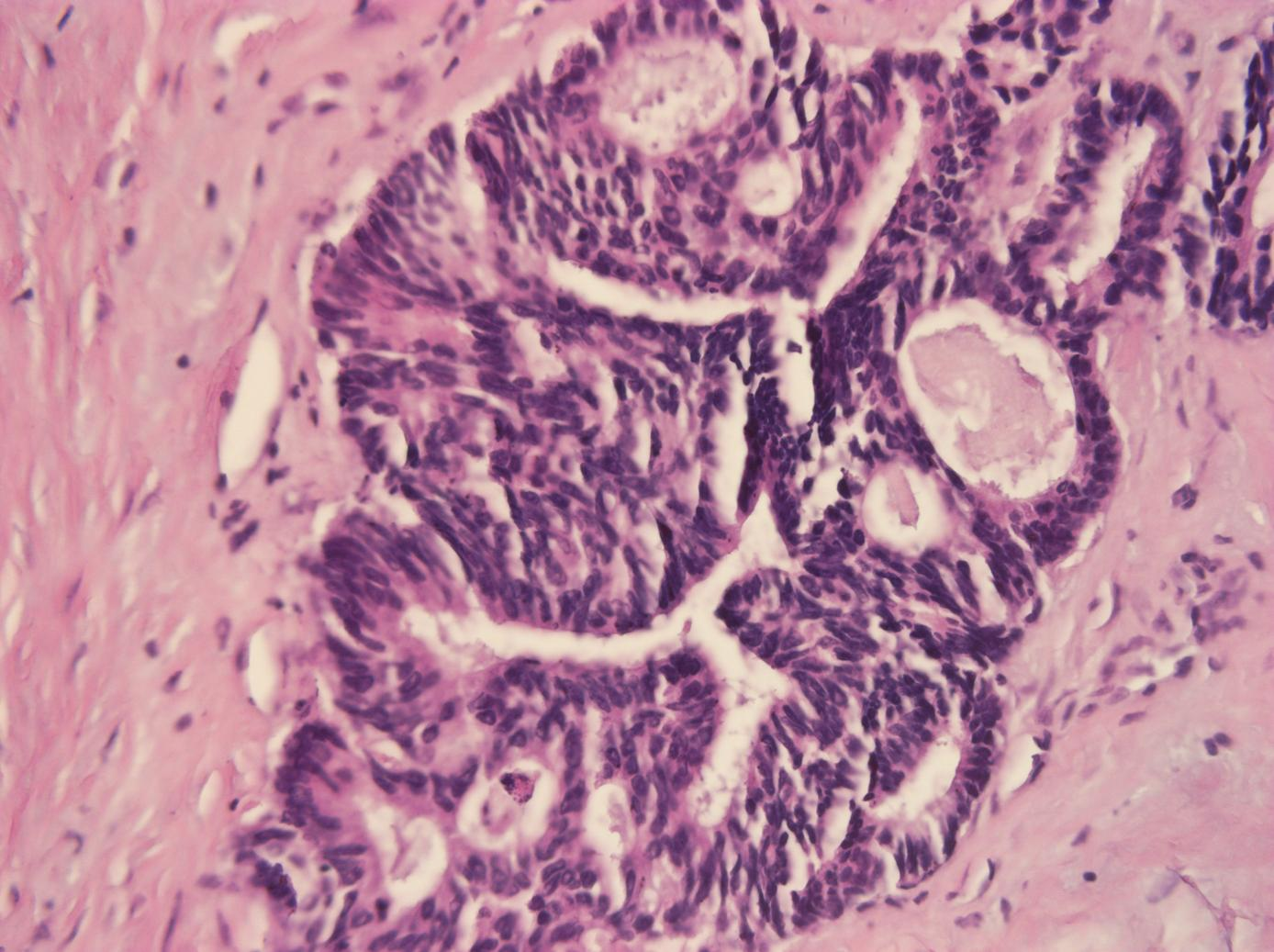}
         \caption{}
         \label{figure:subplot1}
     \end{subfigure}
     \hfill
     \begin{subfigure}[b]{0.24\textwidth}
         \centering
         \includegraphics[width=\textwidth]{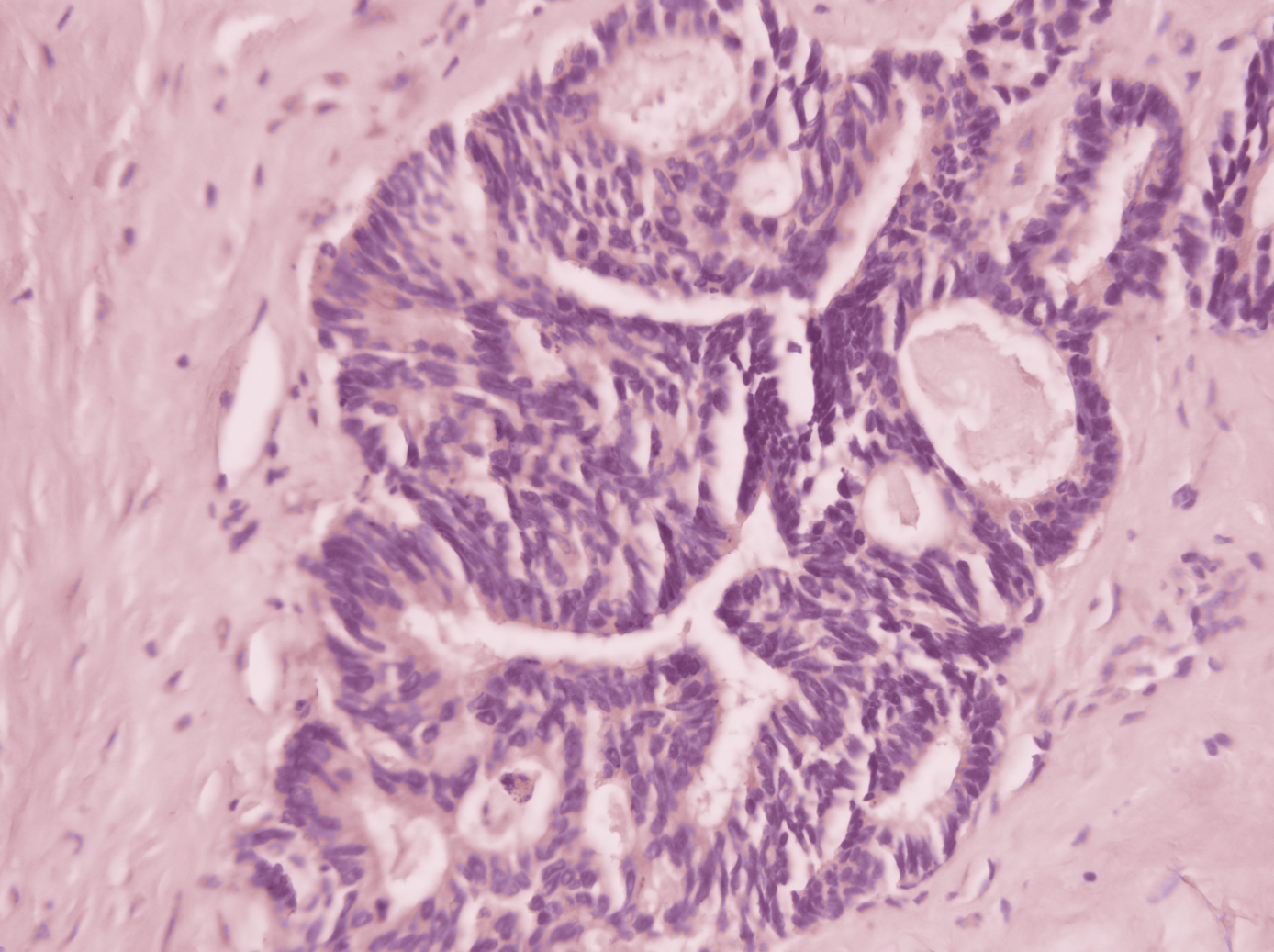}
         \caption{}
         \label{figure:subplot2}
     \end{subfigure}
     
    \bigskip
    
    \begin{subfigure}[b]{0.24\textwidth}
         \centering
         \includegraphics[width=\textwidth]{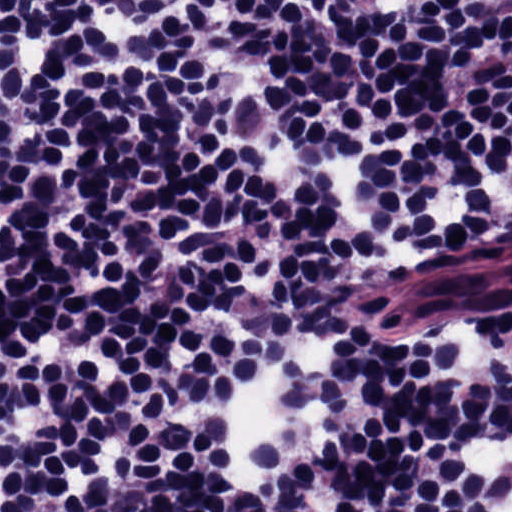}
         \caption{}
         \label{figure:subplot3}
     \end{subfigure}
     \hfill
     \begin{subfigure}[b]{0.24\textwidth}
         \centering
         \includegraphics[width=\textwidth]{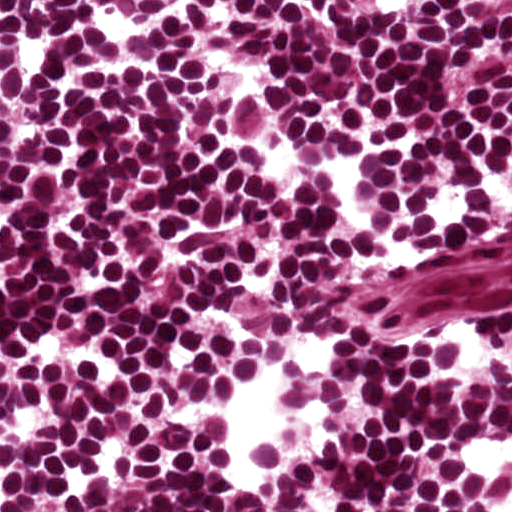}
         \caption{}
         \label{figure:subplot4}
     \end{subfigure}
     \caption{Visualization of raw and stain normalized images. Subfigures (a) and (c) show original images from CoCaHis and CryoNuSeg, respectively. Their stain-normalized equivalents are shown in subfigures (b) and (d).}
     \label{figure:panel}
\end{figure}

\begin{table*}[ht]
\caption{Performance scores of four different models on CryoNuSeg and CoCaHis datasets. Each model had 10 runs with different initial weights. Backbones were initialized on the pretrained ImageNet classification problem. Variable $m\in \{2,3\}$ indicates level of expansion. The best result for each metric and each dataset is in bold font.}
\label{tab:cryonusegPerf}
\resizebox{\textwidth}{!}{%
\begin{tabular}{c|l|cc|cc|cc|}
\cline{2-8}
\multicolumn{1}{l|}{{\color[HTML]{3D3D3D} \textbf{}}} &
  \multicolumn{1}{c|}{{\color[HTML]{3D3D3D} }} &
  \multicolumn{2}{c|}{{\color[HTML]{3D3D3D} \textbf{Original network}}} &
  \multicolumn{2}{c|}{{\color[HTML]{3D3D3D} \textbf{LEFM-Net m=2}}} &
  \multicolumn{2}{c|}{{\color[HTML]{3D3D3D} \textbf{LEFM-Net m=3}}} \\ \cline{3-8} 
\textbf{} &
  \multicolumn{1}{c|}{\multirow{-2}{*}{{\color[HTML]{3D3D3D} \textbf{Metric}}}} &
  \multicolumn{1}{c|}{\cellcolor[HTML]{EFEFEF}\textbf{CryoNuSeg}} &
  \cellcolor[HTML]{C0C0C0}\textbf{CoCaHis} &
  \multicolumn{1}{c|}{\cellcolor[HTML]{EFEFEF}\textbf{CryoNuSeg}} &
  \cellcolor[HTML]{C0C0C0}\textbf{CoCaHis} &
  \multicolumn{1}{c|}{\cellcolor[HTML]{EFEFEF}\textbf{CryoNuSeg}} &
  \cellcolor[HTML]{C0C0C0}\textbf{CoCaHis} \\ \hline
\multicolumn{1}{|c|}{{\color[HTML]{3D3D3D} }} &
  {\color[HTML]{3D3D3D} BACC} &
  \multicolumn{1}{c|}{\cellcolor[HTML]{EFEFEF}{\color[HTML]{3D3D3D} 87.24 $\pm$ 0.54}} &
  \cellcolor[HTML]{C0C0C0}86.06 $\pm$ 1.60 &
  \multicolumn{1}{c|}{\cellcolor[HTML]{EFEFEF}{\color[HTML]{3D3D3D} 86.99 $\pm$ 0.64}} &
  \cellcolor[HTML]{C0C0C0}87.24 $\pm$ 1.55 &
  \multicolumn{1}{c|}{\cellcolor[HTML]{EFEFEF}{\color[HTML]{3D3D3D} 87.52 $\pm$ 0.74}} &
  \cellcolor[HTML]{C0C0C0}88.19 $\pm$ 0.70 \\ \cline{2-8} 
\multicolumn{1}{|c|}{{\color[HTML]{3D3D3D} }} &
  {\color[HTML]{3D3D3D} F\textsubscript{1}} &
  \multicolumn{1}{c|}{\cellcolor[HTML]{EFEFEF}{\color[HTML]{3D3D3D} 82.34 $\pm$ 0.60}} &
  \cellcolor[HTML]{C0C0C0}80.75 $\pm$ 1.44 &
  \multicolumn{1}{c|}{\cellcolor[HTML]{EFEFEF}{\color[HTML]{3D3D3D} 82.11 $\pm$ 0.67}} &
  \cellcolor[HTML]{C0C0C0}81.89 $\pm$ 1.69 &
  \multicolumn{1}{c|}{\cellcolor[HTML]{EFEFEF}{\color[HTML]{3D3D3D} 82.60 $\pm$ 0.69}} &
  \cellcolor[HTML]{C0C0C0}82.85 $\pm$ 1.38 \\ \cline{2-8} 
\multicolumn{1}{|c|}{\multirow{-3}{*}{{\color[HTML]{3D3D3D} \textbf{DeepLabv3+}}}} &
  {\color[HTML]{3D3D3D} PREC} &
  \multicolumn{1}{c|}{\cellcolor[HTML]{EFEFEF}{\color[HTML]{3D3D3D} 85.49 $\pm$ 0.39}} &
  \cellcolor[HTML]{C0C0C0}84.48 $\pm$ 4.70 &
  \multicolumn{1}{c|}{\cellcolor[HTML]{EFEFEF}{\color[HTML]{3D3D3D} 85.74 $\pm$ 0.60}} &
  \cellcolor[HTML]{C0C0C0}83.29 $\pm$ 2.96 &
  \multicolumn{1}{c|}{\cellcolor[HTML]{EFEFEF}{\color[HTML]{3D3D3D} 85.29 $\pm$ 0.79}} &
  \cellcolor[HTML]{C0C0C0}82.76 $\pm$ 3.96 \\ \hline
\multicolumn{1}{|c|}{} &
  {\color[HTML]{3D3D3D} BACC} &
  \multicolumn{1}{c|}{\cellcolor[HTML]{EFEFEF}88.99 $\pm$ 0.87} &
  \cellcolor[HTML]{C0C0C0}86.55 $\pm$ 1.46 &
  \multicolumn{1}{c|}{\cellcolor[HTML]{EFEFEF}89.05 $\pm$ 0.95} &
  \cellcolor[HTML]{C0C0C0}85.58 $\pm$ 2.32 &
  \multicolumn{1}{c|}{\cellcolor[HTML]{EFEFEF}88.73 $\pm$ 0.95} &
  \cellcolor[HTML]{C0C0C0}88.67 $\pm$ 1.64 \\ \cline{2-8} 
\multicolumn{1}{|c|}{} &
  {\color[HTML]{3D3D3D} F\textsubscript{1}} &
  \multicolumn{1}{c|}{\cellcolor[HTML]{EFEFEF}84.95 $\pm$ 0.82} &
  \cellcolor[HTML]{C0C0C0}79.94 $\pm$ 1.70 &
  \multicolumn{1}{c|}{\cellcolor[HTML]{EFEFEF}85.02 $\pm$ 0.87} &
  \cellcolor[HTML]{C0C0C0}78.53 $\pm$ 3.72 &
  \multicolumn{1}{c|}{\cellcolor[HTML]{EFEFEF}84.79 $\pm$ 0.95} &
  \cellcolor[HTML]{C0C0C0}82.59 $\pm$ 1.67 \\ \cline{2-8} 
\multicolumn{1}{|c|}{\multirow{-3}{*}{\textbf{UNet}}} &
  {\color[HTML]{3D3D3D} PREC} &
  \multicolumn{1}{c|}{\cellcolor[HTML]{EFEFEF}88.12 $\pm$ 0.74} &
  \cellcolor[HTML]{C0C0C0}78.60 $\pm$ 5.71 &
  \multicolumn{1}{c|}{\cellcolor[HTML]{EFEFEF}88.17 $\pm$ 1.37} &
  \cellcolor[HTML]{C0C0C0}77.35 $\pm$ 9.59 &
  \multicolumn{1}{c|}{\cellcolor[HTML]{EFEFEF}\textbf{88.73 $\pm$ 1.28}} &
  \cellcolor[HTML]{C0C0C0}80.08 $\pm$ 4.19 \\ \hline
\multicolumn{1}{|c|}{} &
  {\color[HTML]{3D3D3D} BACC} &
  \multicolumn{1}{c|}{\cellcolor[HTML]{EFEFEF}88.98 $\pm$ 0.57} &
  \cellcolor[HTML]{C0C0C0}82.01 $\pm$ 6.57 &
  \multicolumn{1}{c|}{\cellcolor[HTML]{EFEFEF}89.29 $\pm$ 0.60} &
  \cellcolor[HTML]{C0C0C0}86.99 $\pm$ 4.66 &
  \multicolumn{1}{c|}{\cellcolor[HTML]{EFEFEF}88.88 $\pm$ 0.56} &
  \cellcolor[HTML]{C0C0C0}88.23 $\pm$ 2.13 \\ \cline{2-8} 
\multicolumn{1}{|c|}{} &
  {\color[HTML]{3D3D3D} F\textsubscript{1}} &
  \multicolumn{1}{c|}{\cellcolor[HTML]{EFEFEF}84.94 $\pm$ 0.52} &
  \cellcolor[HTML]{C0C0C0}74.04 $\pm$ 9.71 &
  \multicolumn{1}{c|}{\cellcolor[HTML]{EFEFEF}85.27 $\pm$ 0.61} &
  \cellcolor[HTML]{C0C0C0}79.94 $\pm$ 7.37 &
  \multicolumn{1}{c|}{\cellcolor[HTML]{EFEFEF}84.93 $\pm$ 0.61} &
  \cellcolor[HTML]{C0C0C0}81.55 $\pm$ 4.38 \\ \cline{2-8} 
\multicolumn{1}{|c|}{\multirow{-3}{*}{\textbf{UNet++}}} &
  {\color[HTML]{3D3D3D} PREC} &
  \multicolumn{1}{c|}{\cellcolor[HTML]{EFEFEF}88.09 $\pm$ 0.72} &
  \cellcolor[HTML]{C0C0C0}75.64 $\pm$ 17.35 &
  \multicolumn{1}{c|}{\cellcolor[HTML]{EFEFEF}88.04 $\pm$ 0.57} &
  \cellcolor[HTML]{C0C0C0}76.69 $\pm$ 11.94 &
  \multicolumn{1}{c|}{\cellcolor[HTML]{EFEFEF}88.48 $\pm$ 1.08} &
  \cellcolor[HTML]{C0C0C0}78.23 $\pm$ 10.51 \\ \hline
\multicolumn{1}{|c|}{} &
  {\color[HTML]{3D3D3D} BACC} &
  \multicolumn{1}{c|}{\cellcolor[HTML]{EFEFEF}89.30 $\pm$ 0.44} &
  \cellcolor[HTML]{C0C0C0}88.02 $\pm$ 1.22 &
  \multicolumn{1}{c|}{\cellcolor[HTML]{EFEFEF}89.31 $\pm$ 0.38} &
  \cellcolor[HTML]{C0C0C0}88.69 $\pm$ 1.63 &
  \multicolumn{1}{c|}{\cellcolor[HTML]{EFEFEF}\textbf{89.41 $\pm$ 0.29}} &
  \cellcolor[HTML]{C0C0C0}\textbf{89.36 $\pm$ 1.27} \\ \cline{2-8} 
\multicolumn{1}{|c|}{} &
  {\color[HTML]{3D3D3D} F\textsubscript{1}} &
  \multicolumn{1}{c|}{\cellcolor[HTML]{EFEFEF}85.12 $\pm$ 0.51} &
  \cellcolor[HTML]{C0C0C0}82.75 $\pm$ 1.10 &
  \multicolumn{1}{c|}{\cellcolor[HTML]{EFEFEF}85.27 $\pm$ 0.39} &
  \cellcolor[HTML]{C0C0C0}84.33 $\pm$ 1.98 &
  \multicolumn{1}{c|}{\cellcolor[HTML]{EFEFEF}\textbf{85.35 $\pm$ 0.25}} &
  \cellcolor[HTML]{C0C0C0}\textbf{84.96 $\pm$ 1.14} \\ \cline{2-8} 
\multicolumn{1}{|c|}{\multirow{-3}{*}{\textbf{MA-net}}} &
  {\color[HTML]{3D3D3D} PREC} &
  \multicolumn{1}{c|}{\cellcolor[HTML]{EFEFEF}87.36 $\pm$ 0.58} &
  \cellcolor[HTML]{C0C0C0}83.16 $\pm$ 2.44 &
  \multicolumn{1}{c|}{\cellcolor[HTML]{EFEFEF}87.93 $\pm$ 0.46} &
  \cellcolor[HTML]{C0C0C0}\textbf{86.52 $\pm$ 1.60} &
  \multicolumn{1}{c|}{\cellcolor[HTML]{EFEFEF}87.76 $\pm$ 0.53} &
  \cellcolor[HTML]{C0C0C0}86.15 $\pm$ 2.00 \\ \hline
\end{tabular}%
}
\end{table*}

\subsection{Implementation with training and test details}
\label{sec:implDetail}
LEFM layer is implemented in Python using the Pytorch framework \cite{paszke2019pytorch}. All existing deep networks are inherited and modified from \textit{segmentation-models-pytorch} library \cite{yakubovskiy2019segmentationmodels}. For faster convergence \cite{sitnik2021cocahis}, weights of backbones are initialized on ImageNet classification problem \cite{deng2009imagenet} and then fine-tuned with the rest of the LEFM-Net.

Before the training, images were preprocessed as described in Section \ref{sec:preprocessing}. For the evaluation of the statistical significance, all models were trained with 10 different random seeds. Thus, the initialization of newly added weights, coefficients, and choice of patches withheld for validation set were different in every run. In Table \ref{tab:cryonusegPerf}, we show the average performance of models on selected metrics (see Section \ref{sec:metrics}). For each dataset, the corresponding test sets are the same for every model. 

We selected Adam \cite{kingma2014adam} optimizer for the training procedure. To equally punish the underperformance in terms of false positives and false negatives, we used Dice loss function \cite{seo2021closing,eelbode2020optimization}. The hyperparameters for every model were set as follows: 
\begin{itemize}
    \item maximum number of epochs equals to $3\cdot 10^4$
    \item starting learning rate was set to $10^{-3}$
    \item learning rate is reduced when the loss plateau is reached
    \begin{itemize}
        \item waiting patience for improvement is set to 20 epochs
        \item reduction factor is set to 0.5
        \item minimum learning rate equals $10^{-6}$
        
    \end{itemize}
    \item factor for $L_2$ weight decay is set to $10^{-4}$
    \item Early stopping after 40 epochs when there is no improvement on the validation set
    \item data augmentation
    \begin{itemize}
        \item shift limit = 0.2
        \item scale limit = 0.2
        \item rotation limit = 30$^{\circ}$
        \item horizontal and vertical flip
        \item probability = 0.5
    \end{itemize}
    \item LEFM order $m\in\{2,3\}$
\end{itemize}

\subsection{Results}
\label{sec:results}
Results of exhaustive training and testing of four different models, with and without the LEFM layer, on two datasets and 10 random seeds are shown in Table \ref{tab:cryonusegPerf}. Models are ordered by the number of parameters. As mentioned in Section \ref{subsec:dnn}, DeepLabv3+ has the least and MA-net the most trainable weights. When comparing these two models, a noticeable increase in scores can always be found in the model with larger capacity, i.e. MA-net.

It can be seen that four different versions of LEFM-Nets, compared to the original networks, yield better scores on two datasets. Higher order features are important for better end results. As it can be seen in Table \ref{tab:cryonusegPerf}, regardless of the level of expansion, LEFM-Net outperforms original deep models. Also, nearly half of the observed metrics for CryoNuSeg dataset and almost all of them for CoCaHis dataset are better when the level of expansion is $m=3$.

Tests performed on CryoNuSeg show that all LEFM-Nets are on average better than the original network. ANOVA statistical testing indicates that, with statistical significance $\alpha = 0.05$, all LEFM-Nets significantly outperform original models on CoCaHis dataset for $m=3$. Tests are performed on $F_1$ and BACC metrics. Hence, introduction of the LEFM layer seems to be justified for intraoperative pixel-based diagnostics with deep learning models. 

The training time depends on the number of training parameters in the model, dataset, batch- and patch size, and on the degree of expansion $m$. With NVIDIA Tesla V100S GPU, training time for one run ranges from 2 hours, when training DeepLabv3+, to 20 hours when training MA-net.
\section{Discussion and Conclusion}
\label{sec:discAndCon}

\begin{figure*}[ht]
\centering
\includegraphics[width=\textwidth]{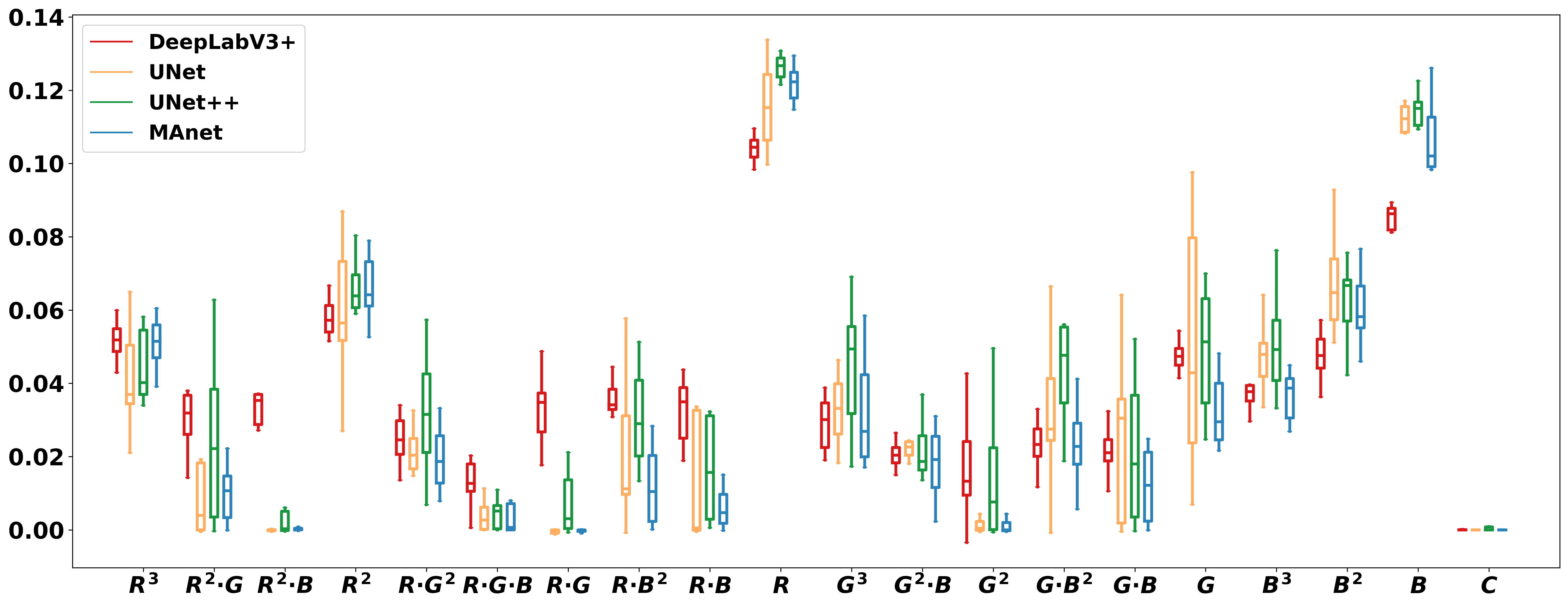}
\caption{Visualization of coefficients learned during the training on CryoNuSeg dataset. DeepLabv3+, UNet, UNet++, and MA-net were trained with expansion layer of degree $m=3$. During the training, $L_2$ norm was used for weight decay. Red ($R$), green ($G$) and blue ($B$) stand for $x_1$, $x_2$ and $x_3$ features, respectively.}
\label{figure:coefficientsD3}
\end{figure*}

The core idea behind the LEFM layer is generation of more explicit features in a data-driven manner, i.e. letting the network to decide on their importance for the image segmentation problem at hand. Towards that, we illustrate in Fig. \ref{figure:coefficientsD3} the relative importance of explicit features induced by the LEFM or order $m=3$ on the CryNuSeg dataset. For the sake of interpretability, we name the abstract features in the input space $\{x_1, x_2, x_3\}$ by the colors they represent \{red, green, blue\}. To interpret Fig. \ref{figure:coefficientsD3} it is important to notice that in images of H\&E stained specimens, nucleus, cytoplasm and glandular structures appear respectively blue-purple, pink and white. Intensely red structures may also represent red blood cells \cite{kothari2014removing}. Since red (R) and blue (B) colors are dominant in the purple color formation, they are dominantly expressed in Fig. \ref{figure:coefficientsD3}. However, justification for using the LEFM layer is because some explicit features, that do not exist in the input data space (monomials of the original features), are also important when the model is segmenting the image. As seen in Fig. \ref{figure:coefficientsD3}, the networks are using them up to certain extents. Hence, when the new features were added, i.e. number of channels is increased, LEFM-Net is expected to be guided towards better predictions while training. Thus, the level of expansion $m$ is a hyperparameter for cross-validation. However, there is a memory limitation to what degree can features be expanded (see Eq. (\ref{eq:binom})). Thus, we limited our experiments to the $m\in \{2,3\}$.

Interpretability of features in the domain of histopathological pixel-wise diagnostics is straightforward. Newly added features are nonlinear mixtures of the original features from the input space. Therefore, they can be interpreted as mixtures of colors needed for discriminating between cancer and non-cancer pixels. Using the previously explained coloring for nuclei, cytoplasm and glandular structures, it is clear that blue-purple, pink and white colors cannot be explained in terms of $R$, $G$ and $B$ features only. Thus, networks also used $RG$, $RB$ and $GB$ features, as seen in Fig. \ref{figure:coefficientsD3}. Since those features do not exist in the input data space, the LEFM layer helps the original network by providing the new medically meaningful features.

It is important to mention that feature expansion brings slightly increased computational complexity as there are more channels to the input of existing neural networks. Therefore, besides newly introduced coefficients in the LEFM layer, new neurons are added in every existing network to adapt for the expanded input. Thus, specific number of new learnable weights depends on the network's architecture. However, as discussed in  Section IV.B the increase is modest.

To conclude, this work introduced a data-adaptive and learnable explicit feature map layer. Besides the features from the input data space, it generates higher order features and helps the existing network to improve segmentation/classification performance. New features are predefined by order of expansion and their importance is learned through the training process. Learned coefficients identify whether the feature is important for classification. Overall, if the original features are interpretable, then their monomials should be interpretable as well. The same applies to color mixtures of H\&E stains in the domain of histopathological image segmentation. By performing exhaustive validation on two different datasets and across multiple runs, introducing the LEFM layer to existing deep models seems to be justified. Presented statistical significance analysis further supports this statement.

\bibliographystyle{IEEEtran}
\bibliography{tmi}


\end{document}